\documentclass[prl,aps,tightenlines,twocolumn,nofootinbib,preprintnumbers]{revtex4}
\begin{document}

\title{Seed perturbations for primordial magnetic fields from MSSM flat
  directions}
\author{Kari Enqvist~$^{1}$, Asko Jokinen~$^{2}$, and Anupam Mazumdar~$^{3}$}
\affiliation{$^{1}$~Physics Department, University of Helsinki, and
Helsinki Institute of Physics, P.O. Box 9, FIN-00014 University of
Helsinki, Finland\\
$^{2}$~Physics Department, University of Helsinki,
P.O. Box 9, FIN-00014 University of Helsinki, Finland\\
$^{3}$~CHEP, McGill University, Montr\'eal, QC, H3A~2T8, Canada.}

\begin{abstract}
We demonstrate that the MSSM flat directions can naturally account for
the seed magnetic fields in the early Universe. The non-zero vacuum
expectation value of an MSSM flat direction condensate provides masses
to the gauge fields and thereby breaks conformal invariance. During
inflation the condensate receives spatial perturbations and
$SU(2)\times U(1)_Y$ gauge currents are generated together with
(hyper)magnetic fields. When these long wavelength vector
perturbations reenter our horizon they give rise to $U(1)_{\rm em}$
magnetic fields with an amplitude of $10^{-30}$~Gauss, as required by
the dynamo mechanism.
\end{abstract}

\preprint{HIP-2004-20/TH}

\maketitle


A large scale magnetic field with a coherence scale of few kpc and
an amplitude of order $\mu G $ is observed today in galaxies and
galactic clusters~\cite{Kronberg}. The origin of the cosmic
magnetic field may well be found in the small primordial magnetic
fluctuations in the early Universe, which are amplified by the
galactic dynamo mechanism~\cite{Kronberg,Beck}. While various
astrophysical processes, such as battery~\cite{Kulsrud} or
vorticity effects~\cite{Harrison} face an uphill task in
sustaining the large scale coherence length~\cite{Grasso}, there
is no shortage of microphysical processes in the early Universe
that could generate a magnetic field~\cite{kenq}. The problem is
usually the shortness of the magnetic coherence length, which however could
be enhanced by the inverse cascade of magnetohydrodynamic
turbulence ~\cite{cascade}.


It has been noted already many years ago that the coherence length
problem might be solved by inflation provided conformal invariance
gets broken during inflation. Interesting possibilities have been
explored  where the conformal invariance is broken either via
triangle-anomaly of the gauge field stress energy tensor~\cite{Dolgov},
or via the gravitational or anomalous coupling of the photon to the
axion~\cite{Widrow}. There are also proposals for exciting the massive
$Z$-boson of the Standard Model during inflation and preheating, which
could produce $U(1)_{\rm em}$ magnetic field after the electroweak
phase transition via hypercharge mixing~\cite{Prokopec}.

In this paper we will discuss a simple mechanism for generating a seed
magnetic field during inflation in the context of the Minimally Supersymmetric
Standard Model (MSSM). As is well known, MSSM has flat directions,
made up of gauge invariant combinations of squarks and sleptons, which
may acquire non-vanishing vacuum expectation values (vevs) during
inflation, thereby forming homogeneous zero-mode condensates. The
condensates may play a significant role in many cosmological
phenomena~\cite{Enqvist}, such as generating baryons and dark matter particles
by first fragmenting into $Q$-balls which then decay \cite{qball}; it
has also been suggested that the origin of all matter and density
perturbations could, in principle, be due to MSSM flat directions
\cite{denspert}.

By definition, flat directions are the minimum energy
configurations even when supersymmetry is broken. Because of the
condensate the SM gauge fields obtain non-vanishing mass terms,
thereby breaking the conformal invariance. This is not by itself
sufficient for generation of magnetic fields, since for the
homogenous condensate the gauge field configurations are pure
gauge~\cite{Asko}. However, during inflation the flat directions
obtain quantum fluctuations from the inflaton induced false
vacuum, which perturbs the classical condensate and also induces
fluctuations in the gauge degrees of freedom. These may no longer
be gauged away. The situation may be compared with plasma physics,
where an electromagnetic field is generated in an electrically
neutral plasma by virtue of the motion of the electrons. Here the
condensate perturbations represent also perturbations in the
charge density and by virtue of their motions, generate gauge
currents.

While inflation lasts, the gauge field perturbations are stretched
outside the horizon. When they eventually re-enter they provide us
with a seed (hyper)magnetic field, which after electroweak phase
transition is projected onto the regular $U(1)_{\rm em}$ magnetic
field.

Let us consider one particular flat direction with components
carrying $SU(2)\times U(1)_{Y}$ charges such as $LL\bar e$. The
field content of this direction  is
given by
\begin{equation}\label{LLe}
\label{field2.9}
L_i = \frac{\phi}{\sqrt{3}}
\left(\begin{array}{ll}
1 \\ 0 \end{array}\right), ~~
L_j =\frac{\phi}{\sqrt{3}}\left( \begin{array}{ll} 0 \\ 1 \end{array}\right),
~~ \bar{e}_k =\frac{\phi}{\sqrt{3}}
\end{equation}
where $i\neq j$, $\phi$ is a complex scalar field and square roots
are to obtain canonical normalization. The background gauge fields are assumed to vanish. With the field
configuration Eq.~(\ref{field2.9}) it is possible to show that the
gauge invariant matter currents
\begin{eqnarray}
\label{curr2.6}
J_{\mu}^A &=& ig \sum_i \left[\phi_i^{\dagger} T^A D_{\mu}\phi_i - (D_{\mu}
  \phi_i)^{\dagger} T^A \phi_i \right]\,,  \nonumber \\
J_{\mu}^Y &=& ig' \sum_i Y_i \left[\phi_i^{\dagger} D_{\mu} \phi_i - (D_{\mu}
  \phi_i)^{\dagger} \phi_i \right]\,,
\end{eqnarray}
vanish for vanishing gauge fields in a Robertson-Walker background
$g_{\mu\nu}=a^2(\eta)\eta_{\mu\nu}$, where $a(\eta)$ is the scale
factor in comoving coordinates and $\eta$ the conformal time.  The covariant
derivative is given by $D_{\mu} = \partial_{\mu} - ig A_{\mu}^A T^A - ig'Y
A_{\mu}$, where $A_{\mu}(A^{A}_{\mu})$ is the Abelian (non-Abelian) gauge
field corresponding to $U(1)_{Y}$ ($SU(2)$). The summation in
Eq.~(\ref{curr2.6}) is over the scalar fields belonging to the
representation of the corresponding gauge group.

The corresponding equations of motion are given by
\begin{eqnarray}
\label{eqm2.4}
\eta^{\rho\nu} D_{\rho}^{(Adj)} F_{\mu\nu}^A = a(\eta)^2 J_{\mu}^A\,,
\nonumber \\
\eta^{\rho\nu} \partial_{\rho} F_{\mu\nu} = a(\eta)^2 J_{\mu}^Y\,,
\nonumber \\
\eta^{\mu\nu} D_{\mu} D_{\nu} \phi_i + 2\mathcal{H} D_0 \phi_i +
\frac{\partial V}{\partial \phi_i^{\dagger}} = 0\,.
\end{eqnarray}
Here $\mathcal{H}$ is the comoving Hubble expansion rate, and the
adjoint covariant derivative is determined by $D_{\rho}^{(Adj)}
F_{\mu\nu}^A = \partial_{\rho} F_{\mu\nu}^A + g f^{ABC}A_{\rho}^B
F_{\mu\nu}^C$, where $f^{ABC}$ is the structure constant of the
$SU(2)$ gauge group, and $F_{\mu\nu}^A$ is the gauge field
strength.

Perturbing Eq.~(\ref{eqm2.4}) to first order we get\footnote{We have
neglected the perturbations of metric and non-flat degrees of
freedom. In the gauge $\delta A_0^A=0$ and $\partial^i \delta A_i^A=0$
the result for the gauge field equation turns out to be the same.}
\begin{eqnarray}
\label{pert2.10}
\eta^{\rho\nu} \partial_{\rho} \left(\partial_{\mu} \delta A_{\nu}^A -
  \partial_{\nu} \delta A_{\mu}^A \right) = a^2 \delta J_{\mu}^A \nonumber \\
\eta^{\rho\nu} \partial_{\rho} \left(\partial_{\mu} \delta A_{\nu} -
  \partial_{\nu} \delta A_{\mu} \right) = a^2 \delta J_{\mu}^Y
\end{eqnarray}
where the background equation $A_{\mu}^A=A_{\mu}=0$ has been used. In
calculating the perturbations of the matter currents the
perturbations of the flat direction, $\delta\phi$, do not appear
since the current vanishes for all values of $\phi$ for vanishing
gauge fields, so only the terms proportional to gauge field
perturbations contribute:
\begin{eqnarray}
\label{pert2.11}
\delta J_{\mu}^A &=& ig \left[L_i^{\dagger} T^A (-ig T^B \delta A_{\mu}^B -
  ig'Y_L \delta A_{\mu}) L_i \right. \nonumber \\
& & \left. - L_i^{\dagger} (ig T^B
  \delta A_{\mu}^B + ig'Y_L \delta A_{\mu}) T^A L_i + (L_i \to L_j) \right]
\nonumber \\
\delta J_{\mu}^Y &=& ig' \left[2 Y_L L_i^{\dagger} (-ig T^B \delta A_{\mu}^B -
  ig'Y_L \delta A_{\mu}) L_i \right. \nonumber \\ & & \left. + (L_i \to L_j) -
  2 ig' Y_{\bar{e}}^2 \delta A_{\mu} |\bar{e}_k|^2 \right]
\end{eqnarray}
Inserting the background field configuration of the flat direction
Eq.~(\ref{field2.9}) into Eq.~(\ref{pert2.11}), we finally obtain
\begin{eqnarray}
\label{pert2.12}
\delta J_{\mu}^A &=& \frac{1}{3} g^2 |\phi|^2 \delta A_{\mu}^A \nonumber \\
\delta J_{\mu}^Y &=& g'^2 |\phi|^2 \delta A_{\mu}
\end{eqnarray}
Hence Eq.~(\ref{pert2.12}) produces effective mass terms for
Eq.~(\ref{pert2.10}). The equations of motion are similar to
massive scalar field equations of motion and thus in the Coulomb
gauge $\delta A_0^A = \delta A_0 = \partial^i \delta A_i^A =
\partial^i \delta A_i = 0$, Eq.~(\ref{pert2.10}) may be written as
\begin{eqnarray}
\label{pert2.13}
(\partial_{\eta}^2 - \partial^j \partial_j) \delta A_i^A +
\frac{1}{3} g^2 |\phi|^2 a(\eta)^2 \delta A_i^A &=& 0 \nonumber \\
(\partial_{\eta}^2 - \partial^j \partial_j) \delta A_i + g'^2
|\phi|^2 a(\eta)^2 \delta A_i &=& 0\,,
\end{eqnarray}
where $i=1,2,3$. Eq.~(\ref{pert2.13}) reminds us of a scalar field
equation with mass $m=|g\phi|/\sqrt{3}$ and $m'=|g'\phi|$. The
solutions can be written as plane wave expansions
\begin{eqnarray}
\label{expansiona}
\delta A_i^A = \sum_{\bf k} \left( w_k(\eta) e_i^{A\lambda} a_{{\bf k}\lambda}
  + w_k(\eta)^* e_i^{A\lambda *} a_{-{\bf k} \lambda}^{\dagger} \right) e^{i
  {\bf k} \cdot {\bf x}} \,,
\end{eqnarray}
where a corresponding expansion holds for $\delta A_i$, too.
Here $e_i^{A\lambda}$ are the polarization vectors which obey $k^i e_i=0$ in the Coulomb gauge.
The mode functions, $w_k(\eta)$, can be given in terms
of Hankel functions as
\begin{eqnarray}
\label{pert2.14}
w_k(\eta) &=& \frac{\sqrt{-\pi\eta}}{2(2\pi)^{3/2}} e^{i\left(\nu +
    \frac{1}{2}\right)\frac{\pi}{2}} H_{\nu}^{(1)} (-k\eta)
\end{eqnarray}
and similar expression holds for the mode function of $\delta A_{i}$ with
$\nu\rightarrow \nu^{\prime}$. In the above expression $k$ is the wavenumber
and
\begin{equation}
\nu^2 = \frac{1}{4} - \frac{g^2|\phi|^2}{3H_I^2}\,,
\quad \nu'^2 = \frac{1}{4} -\frac{g'^2|\phi|^2}{H_I^2}\,,
\end{equation}
where $H_{I}$ is the Hubble expansion rate during inflation. The mode
functions (\ref{pert2.14}) reduce to the vacuum mode functions in the
limit $\eta\to -\infty$,
\begin{equation}
\label{vacuum}
w_k^{(vac)}(\eta) = \frac{1}{(2\pi)^{3/2} \sqrt{2k}}\, e^{-ik\eta}
\end{equation}
We are mainly interested in the super Hubble horizon fluctuations
of $\delta A_{i},\delta A_{i}^{A}$. When $k|\eta|\ll 1$, we obtain,
\begin{eqnarray}
\label{perta}
\left|\delta A^{A}_{i}\right|^2 &\approx & \frac{2|\Gamma(\nu)|^2}{(2\pi)^4 k}
\left(\frac{k}{2H_{I}}\right)^{1-2\nu} \,.
\end{eqnarray}
and a similar expression for $\left|\delta A_{i}\right|^2$ but with
$\nu\rightarrow \nu^{\prime}$.  The amplitudes of the long wavelength
fluctuations are frozen and start to re-enter once inflation comes to
an end. For the sake of simplicity we assume here that
post-inflationary era does not trigger super-Hubble fluctuations in
the gauge fields which would alter our result in
Eq.~(\ref{perta}). However, note that the vacuum structure is
different during inflation and the radiation era\footnote{Note that
the flat direction condensate amplitude slides both during inflation
and post-inflationary era.  Therefore the masses $m,m^{\prime}$ vary
with time. The MSSM flat direction decays late but well before the
electroweak scale \cite{Marieke}, and by decaying they restore the
conformal invariance.}. Therefore at late times it is important to
match the above mode functions (valid during inflation) to a linear
combination of traveling waves in the radiation dominated epoch via
Bugoliubov coefficients, $\alpha_{k},~\beta_{k}$,
\begin{eqnarray}
w_k(\eta) &=& \alpha_{k} w_k^{(vac)}(\eta) + \beta_{-k}^{\ast}
w_k^{(vac)}(\eta)^* \,, \nonumber \\
\partial_{\eta} w_k(\eta) &=&\alpha_{k}\partial_{\eta} w_k^{(vac)}(\eta) +
\beta^{\ast}_{-k}\partial_{\eta} w_k^{(vac)}(\eta) \,,
\end{eqnarray}
For simplicity let us assume that the transition from inflation to radiation
happens instantly, and we also assume that MSSM flat direction vev becomes
zero right after the end of inflation so that the condensate disappears instantaneously.
Although the MSSM flat direction vev
always goes to zero, it need not vanish right away after the end of
inflation. However, it is possible to find an example within D-term
inflationary models where the flat direction vev can become zero right
after the end of inflation; this happens due to a large Hubble induced mass correction
which causes the condensate field to roll down to the origin within one Hubble
time~\cite{Kolda}.

In this letter we proceed with the assumption of a sudden transition
from inflation to radiation on super horizon scales, which leads to
\begin{eqnarray}
|\beta_{k}|^2 &\approx& \frac{|\Gamma(\nu^{\prime})|^2}{16\pi}
\left(\frac{1}{2}-\nu^{\prime}\right)^2\left(\frac{2H_{I}}{k}
\right)^{1+2\nu^{\prime}}\,,
\end{eqnarray}
and a similar expression for $|\beta_{k}^{A}|^2$ with
$\nu^{\prime}\rightarrow \nu$. Therefore during the radiation era the
amplitude of the gauge fields is enhanced by a factor $\beta_{k}$ compared to the Minkowski vacuum. Note that
during radiation era, $w_k^{(vac)}(\eta) \propto \sqrt{1/2k}e^{-ik\eta}$.
Therefore the final the gauge field spectrum $\propto k^{-1-\nu}$. For
$\nu\approx 1/2$, the spectrum is indeed flat, similar to the case of massless scalar
field. Similar conclusions were also reached in Ref.~\cite{Prokopec}.

During radiation era, at scales where diffusion can be neglected, the
hypercharge field is  frozen in the sense that the lines of
force move together with the fluid, $A_i,~A^A_{i}\propto a(t)^{-2}$.

In our case, the hypercharge field is a linear combination of the
fields, $Y_{\mu}\sim {\cal B}_{\mu} + Z_{\mu}$. We are mainly
concerned with the $U(1)_{\rm em}$ photon field, ${\cal B}_{\mu}$,
below the electroweak scale, given by
${\cal B}_{\mu}=\sin\theta_{W} A_{\mu}^{3}+\cos\theta_{W}A_{\mu}$,
where $\sin\theta_{W}\approx 0.23$ is the Weinberg's angle. Since the
non-Abelian field $A_{\mu}^A$ obtains a screening mass in a radiation
dominated Universe \cite{Biro}, its contribution vanishes and the
unscreened photon field is ${\cal B}_{i}\approx \cos\theta_{W}
Y_{i}$ and the real magnetic field is $B_i = \epsilon_{ijk} \partial_j {\cal
  B}_k$.

Let us now estimate the amplitude of the magnetic field at the length scale
$l$ where $k=2\pi/l\ll H_{I}$. We obtain
\begin{equation}
\label{magnetic1}
B_{l} \approx \cos\theta_{W}~C(\nu')\left(\frac{1}{2}-\nu'\right)H_{I}^2
(l~H_{I})^{\nu'-3/2}~,
\end{equation}
where
\begin{equation}
C(\nu') = 3\cdot 2^{2\nu'} \Gamma(\nu')
\sqrt{\frac{2\Gamma(1-\nu')}{\sqrt{\pi}\Gamma(3/2+\nu')}}\,.
\end{equation}
Within our setup $LL\bar e$ obtains a vanishing vev right after
inflation so that  the amplitude of
the magnetic field is frozen during the radiation epoch, evolving as $B\sim a^{-2}$.
Since the temperature scales as the inverse of the of the scale
factor, the magnetic field at the time of galaxy formation is given by
$B_{l,0}=B_{l}(T_{gf}/T_R)^2$, where $T_{gf}$ is the
temperature at the galaxy formation and $T_{R}$ is the reheat temperature.
The length scale also evolves, $l\sim a$, so the length scale at the galaxy
formation is related to the one at the end of inflation by
$l_{gf} = l(T_R/T_{gf})$. Putting everything together we obtain the
amplitude of the magnetic field at the time of galaxy formation as
\begin{eqnarray}
\label{magnetic2}
B_{l,gf} &\approx &\cos\theta_W C(\nu')
\left(\frac{1}{2} - \nu'\right) H_I^{1/2+\nu'} \, l_{gf}^{\nu' - 3/2}
\,\nonumber \\
&&  \times\left(\frac{T_{gf}}{T_R}\right)^{1/2 + \nu'} \,.
\end{eqnarray}
We estimate the reheat temperature by assuming that the inflaton
potential energy density is directly converted to a thermal bath, $V_{end}\approx
3M_p^2 H_I^2 \approx (\pi^2/30) g_* T_R^4$, where $g_{\ast}=106.75$,
is the number of relativistic degrees of freedom. Thus we obtain an order of
magnitude estimate for the magnetic field which reads
\begin{eqnarray}
\label{magnetic3}
B_{l,gf} &\approx & 7\cdot 10^{-30} \rm{G} \,\,(1+z_{gf})^{-1}
    \left(\frac{g'}{0.01}\right)^2
    \left(\frac{|\phi|}{H_{I}}\right)^2\, \nonumber \\
 && \times \left(\frac{T_0}{2.73~\rm{K}}\right)
    \left(\frac{1~\rm{kpc}}{l_{gf}}\right) \left(\frac{V_{end}^{1/4}}{10^{16}~
    \rm{GeV}}\right)\,,
\end{eqnarray}
when $(1/2 - \nu') \approx (g'|\phi|/H_I)^2 \ll 1/2$ and
$T_{gf}=(1+z_{gf})T_0$, where $z_{gf}$ is the redshift of the galaxy formation
time and a factor of $5\cdot 10^4(1+z_{gf})^{-2}$ \cite{Enqvist:sa} due to the
increase of flux at the collapse of protogalaxy is included.

Note that the the strength of the magnetic field depends
on the vev of the hypermangetic field during inflation. Obviously when the vev
is zero, the conformal invariance is restored and there is no magnetic
field. For a  field almost massless during inflation, we would naturally expect
that the MSSM flat direction vev is locked to the expansion of the Universe
such that $|\phi|\sim H_{I}$. This yields  the maximum amplitude of the
primordial magnetic field to be of order $10^{-30}~G$ with a coherence length
$l \sim 1$ kpc as required in \cite{Lilley}.

A word of caution, though: the above estimation of the
strength does not take into account of the sub-horizon plasma physics
due to charge separation and inhomogeneous current distribution, see
for instance~\cite{Giovannini1}.  Note that MSSM flat directions do
not generate currents at any time but nevertheless the current density obtains
non-vanishing fluctuations, which requires a detailed analysis of the
sub-horizon effects which goes beyond the scope of the present paper. On the
other hand the condensate is assumed to decay instantly at the end of
inflation, so that conformal invariance of the gauge field is restored. In
reality one would expect the condensate to be stuck at its inflationary
value until the Hubble parameter is of the order of its mass $\sim 1$TeV. Thus
the magnetic field amplitude decay rate is slower than in the conformally invariant
case. Also in reality the condensate should be slow-rolling during and
after inflation unless it is at the minimum of the potential, but we expect
that this produces only minor quantitative changes to our results.

In conclusion, we have provided a natural mechanism within
supersymmetry for generating seed magnetic field based on generating
non-vanishing masses to the SM gauge fields from the vev of the MSSM
flat directions. The simplicity of our scenario lies in that we do not
break conformal invariance by an ad-hoc assumption, nor do we invoke any
exotic fields or coupling other than the ones present in the minimal
extension of the SM.
The excited gauge fields, mainly the hypercharge field arising from
$LL\bar e$ flat direction, are stretched during inflation outside
the horizon in a similar way as the scalar fluctuations responsible
for generating the perturbations in the cosmic microwave background
radiation. The subtle point here is that we generate non-vanishing
current density required for breaking the conformal invariance through
the fluctuating current.

K.E. is supported partly by the Academy of Finland grant no. 75065, and
A.M. is a CITA-National fellow and his work is also supported in
part by NSERC (Canada) and by the Fonds de Recherche sur la
Nature et les Technologies du Qu\'ebec.


\vskip20pt


\begin{thebibliography}{100}

\bibitem{Kronberg}
P.~P.~Kronberg,
Rept.\ Prog.\ Phys.\  {\bf 57}, 325 (1994).




\bibitem{Beck}
R.~Beck, A.~Brandenburg, D.~Moss, A.~Shukurov and D.~Sokoloff,
Ann.\ Rev.\ Astron.\ Astrophys.\  {\bf 34}, 155 (1996).

\bibitem{Kulsrud}
R. M. Kulsrud and S. W. Anderson, Astrophys. J. {\bf 396}, 606 (1992).

\bibitem{Harrison}
E. R. Harrison, Nature {\bf 224}, 1089 (1969); Mon. Not. R. Astron. Soc. {\bf 147},
279 (1970); Phys. Rev. Lett. {\bf 30}, 188 (1973).

\bibitem{Grasso}
D.~Grasso and H.~R.~Rubinstein,
Phys.\ Rept.\  {\bf 348}, 163 (2001).
M. Giovannini, astro-ph/0312614.

\bibitem{kenq}
Kari Enqvist, Int.~J.~Mod.~Phys.~{\bf D7}, 331 (1998).

\bibitem{cascade}
A. Brandenburg, K. Enqvist, and P. Olesen, Phys.Rev. D {\bf 54},
1291 (1996); M. Christensson, M. Hindmarsh, and A. Brandenburg,
Phys.Rev. E {\bf 64}:056405 (2001).



\bibitem{Dolgov}
A. D. Dolgov, Zh. Eksp. Teor. Fiz. {\bf 81}, 417
(1981) [Sov. Phys. JETP {\bf 54} (2), 223 (1981)]; Phys. Rev. D {\bf
48}, 2499 (1993).

\bibitem{Widrow}
M.~S.~Turner and L.~M.~Widrow,
Phys.\ Rev.\ D {\bf 37}, 2743 (1988).


\bibitem{Prokopec}
A.~C.~Davis, K.~Dimopoulos, T.~Prokopec and O.~Tornkvist,
Phys.\ Lett.\ B {\bf 501}, 165 (2001);
K.~Dimopoulos, T.~Prokopec, O.~Tornkvist and A.~C.~Davis,
Phys.\ Rev.\ D {\bf 65}, 063505 (2002).


\bibitem{Enqvist}
K.~Enqvist and A.~Mazumdar,
Phys.\ Rept.\  {\bf 380} (2003) 99.

\bibitem{qball}
A. Kusenko, and M. E. Shaposhnikov, Phys. Lett. B {\bf 418}, 46
(1998); G. R. Dvali, A. Kusenko, and M. E. Shaposhnikov, Phys.
Lett. B {\bf 417}, 99 (1998); K.~Enqvist and J.~McDonald, Phys.\
Lett.\ B {\bf 425}, 309 (1998);  K.~Enqvist and J.~McDonald,
Nucl.\ Phys.\ B {\bf 538}, 321 (1999).

\bibitem{denspert}
K.~Enqvist, S.~Kasuya and A.~Mazumdar, Phys. Rev. Lett. {\bf 90}, 091302 (2003);
K.~Enqvist, A.~Jokinen, S.~Kasuya and A.~Mazumdar, \ Rev.\ D {\bf
68}:103507 (2003);
K.~Enqvist, S.~Kasuya and A.~Mazumdar,
hep-ph/0311224;
 K. Enqvist, A. Mazumdar, A. Perez-Lorenzana, hep-th/0403044.

\bibitem{Asko}
K.~Enqvist, A.~Jokinen and A.~Mazumdar,
JCAP {\bf 0401}, 008 (2004).

\bibitem{Marieke}
M.~Postma and A.~Mazumdar,
JCAP {\bf 0401}, 005 (2004).

\bibitem{Kolda}
C. Kolda, J. March-Russell, Phys. Rev. D {\bf 60}, 023504 (1999).


\bibitem{Biro}
T.~S.~Biro and B.~Muller,
Nucl.\ Phys.\ A {\bf 561}, 477  (1993).

\bibitem{Enqvist:sa}
K.~Enqvist, V.~Semikoz, A.~Shukurov and D.~Sokoloff,
Phys.\ Rev.\ D {\bf 48}, 4557 (1993).

\bibitem{Lilley}
A.~C.~Davis, M.~Lilley and O.~Tornkvist,
Phys.\ Rev.\ D {\bf 60}, 021301 (1999).

\bibitem{Giovannini1}
M. Giovannini and M. Shaposhnikov, Phys. Rev. D {\bf 62}, 103512 (2000).

\end{thebibliography}
\end{document}